\documentclass[aip,reprint]{revtex4-1}

\usepackage{graphicx}
\usepackage{multirow}
\usepackage{amsmath,amssymb,amsfonts}
\usepackage{booktabs}

\usepackage{subcaption}
\usepackage{siunitx}

\begin{document}

%\preprint{}

\title[High Temporal and Spatial Resolution X-Ray and Gamma-Ray Imaging using an Event-Mode  Scintillator-Based Detector]{High Temporal and Spatial Resolution X-Ray and Gamma-Ray Imaging using an Event-Mode  Scintillator-Based Detector}

\author{Alexander Wolfertz}
    \email{alexander.wolfertz@frm2.tum.de}
    \affiliation{Forschungs-Neutronenquelle Heinz Maier-Leibnitz, Technical University of Munich, 85748 Garching, Germany}

\author{Ondrej Zapadlik}
    \affiliation{Crytur, Ltd., Na Lukách 2283, Turnov, 511 01, Czech Republic}

\author{Alex Gustschin}
    \affiliation{Chair of Biomedical Physics, Technical University of Munich, 85748 Garching, Germany}

\author{Adrian Losko}
    \affiliation{Forschungs-Neutronenquelle Heinz Maier-Leibnitz, Technical University of Munich, 85748 Garching, Germany}

\author{Andrei Nomerotski}
    \affiliation{Czech Technical University, Prague 11519, Czech Republic}
    \affiliation{Institute of Physics of Czech Academy of Sciences, Prague 18200, Czech Republic}
    \affiliation{Florida International University, Miami FL 33174, USA}

\date{\today}

\begin{abstract}

LumaCam detectors are a novel type of event-mode imaging detectors based on scintillator screens that have been developed recently for neutron imaging applications. They operate in an event-mode in which individual interactions of the incoming particles with the scintillator screen are reconstructed in both space and time, providing simultaneous high spatial and temporal resolution. In this work, we demonstrate the applicability of LumaCam detectors to X-ray and gamma-ray imaging. First tests with a $120$~kV acceleration voltage X-ray source show a spatial detector resolution down to $50$~{\textmu}m and the temporal resolution is estimated to be ${\sim}1$~{\textmu}s. This combination makes LumaCam detectors especially suitable for imaging fast processes with high spatial resolution with limited cone beam magnification. The most significant adaption of the detector from the version for neutron detection is the scintillator screen. To further investigate the potential of LumaCam detectors for high-energy photon applications, additional measurements employing different scintillator screen materials have been carried out using gamma ray sources. The results show significant improvements in the temporal resolution with values as low as ${\sim}15$~ns. In addition, some scintillators also show significant energy sensitivity in the photon multiplicity spectrum, highlighting the potential of LumaCam detectors for energy-selective X-ray and gamma-ray imaging.
\end{abstract}

%\keywords{X-Ray, Gamma, Imaging, Radiography, Event-Mode, LumaCam, Scintillator}


\maketitle

\section{Introduction}\label{sec:introduction}

    The field of position-sensitive X-ray detection has traditionally been dominated by two main technologies: direct detection semiconductor sensors and indirect detection scintillator-based systems. While direct detection offers excellent energy resolution and efficiency at lower energies (${<}20$~keV), it faces significant hurdles at the higher energies required e.g. for medical radiography. A primary challenge is charge sharing and the escape of secondary radiation from the detection volume \cite{https://doi.org/10.1118/1.4820371, Danielsson_2021}. These issues are particularly problematic in detectors with small pixel sizes, which is why photon-counting at microscopic resolution has remained a challenge. Furthermore, current direct conversion detectors struggle with large-area coverage due to sensor stitching limitations, crystal quality issues \cite{https://doi.org/10.1118/1.4820371}, and the high cost and complexity of their readout electronics \cite{Danielsson_2021}.

    An alternative approach is to combine indirect conversion with light detection sensors. Conventional flat-panel detectors, where a scintillator is directly coupled to a Thin-Film Transistor (TFT) photodiode matrix, typically employ pixel sizes in the range of 200 {\textmu}m, which lack the requisite sensitivity and speed to detect single X-ray photons. An emerging technology allowing single photon detection is coupling the scintillator to silicon photomultipliers (SiPMs)~\cite{10.1021/acs.chemmater.4c03437}, with current research into fast scintillators aiming to further improve count-rate and temporal resolution capabilities~\cite{LECOQ2016130}. High-resolution X-ray detectors, driven by developments at synchrotron sources, rely on coupling via intermediate optical devices to highly sensitive sensors with pixel sizes in the sub-10~{\textmu}m range\cite{Martin:gf0006}. The most common approaches are fiber-optic taper (FOT) bonded assemblies and lens-coupled systems \cite{Uesugi:gf5031}. These imaging systems primarily utilize either thin powder-based screens (e.g., Gd$_2$O$_2$S ) or single-crystal scintillators (e.g., GAGG, YAP). While easier to manufacture, powder screens exhibit lower X-ray stopping power due to their grainy structure and reduced packing density. Conversely, single crystals provide higher material density and homogeneous stopping power, though fabricating them into ultra-thin substrates is significantly more complex. Furthermore, their light exit mechanisms differ fundamentally; optical scattering at grain boundaries in powder screens introduces a diffuse background, whereas transparent single crystals eliminate scattering but are heavily constrained by total internal reflection~\cite{Tengattini:22}.

    FOT-based architectures utilize fiber-optic matrices bonded directly to the image sensor. To mitigate background noise in high-flux environments, these tapers frequently incorporate highly absorbing extramural glass structures, which attenuate a significant fraction of the primary beam. Similar to lens coupled alternatives, they can provide image magnification or de-magnification. An advantage of FOT systems compared to lenses is the significantly higher light collection efficiency owing to their superior numerical aperture.

    Lens coupled systems offer distinct advantages regarding experimental versatility. By utilizing adjustable optics and mirrors, they allow for a variable field of view and adaptable spatial resolution. Lens based systems also eliminate the fiber diameters limitations of FOT based systems, and structural artifacts arising from the hexagonal packing and alignment matrix of the fibers. Additionally, the scintillator screen can be effortlessly swapped to meet shifting requirements for detection efficiency, as well as spatial and temporal resolution. Critically, positioning the sensor entirely outside of the primary beam path isolates the sensitive readout electronics from radiation damage and electronic interference, making lens-coupled configurations highly robust in intense radiation environments.
    
    Scintillator screens coupled to CCD sensors have been shown to be capable of energy-resolved X-ray detection with moderate energy resolution while achieving sub-pixel spatial resolution~\cite{1462377}. An FOT-coupled configuration employing photon-counting operation was further described by O'Connell et al.~\cite{OConnell:20}, demonstrating an increased spatial resolution relative to conventional integrating imaging using a $10$~{\textmu}m thick Gd$_2$O$_2$S:Tb screen tapered to the sensor. However, such systems remain fundamentally limited by the frame rate (currently typically $1-10$~ms exposure time at full pixel matrix readout), and inefficient readout mechanisms, which restrict their ability to handle higher radiation fluxes.

    %Scintillators play a crucial role in the optical detection of X-rays and soft gamma rays by converting high-energy radiation into visible light, which can then be captured by an optical camera. This approach was used before to build 2D spatially resolved detectors with slow cameras \cite{Gu2025}. One of the main advantages of the scheme is decoupling of the sensing and detection elements. The optical camera can be placed far away from the scintillator and outside the direct beam. Another advantage is a high flexibility in field-of-view (FoV) and resolution achievable by changing the optical elements.

    LumaCam detectors~\cite{losko2021new, Wolfertz2024} have recently been developed for neutron imaging~\cite{Gustschin2024, MIRASHI2025170284, Hirsh2025, Jager2025}, diffraction~\cite{Jager:in5094}, and reflectometry~\cite{Khaplanov2025} applications, and have also been used for alpha particle detection~\cite{DAmen2021}. They consist of a scintillator screen to convert incident particles to flashes of visible light, an image intensifier to amplify the signal, and an image sensor to capture the intensified light. Similar to the CCD- and CMOS-based photon-counting detectors described above, LumaCam detectors also operate in event-mode. However, the additional image intensifier allows the usage of less light sensitive but significantly faster imaging chips. This not only improves the achievable time resolution ($\lesssim 10$~ns has been demonstrated~\cite{Wolfertz2024_1}) but also makes LumaCam detectors suitable for higher flux applications ($\gtrsim 1$~MHz total count rate). The amount of scintillation light generated and its temporal distribution depends on the scintillator type. Typically, LumaCam detectors capture several photons from the flash of a single particle, measuring their position and time. Detecting multiple photons allows for an enhanced identification of particles including background rejection, enhanced spatial and temporal accuracy, and afterglow suppression.

    Although originally developed for neutrons, LumaCam detectors can also detect X-rays and gamma rays, which was previously observed as a parasitic background contribution during neutron measurements. This work shows data from dedicated X-ray and gamma ray measurements in the energy range from $10$~keV to $660$~keV, exploring the potential for LumaCam detectors in the field of space- and time-resolved high-energy photon detection.

    The most significant difference in the detector physics of a LumaCam between neutron and high-energy photon detection are the interactions in the scintillator. Scintillator screens optimized for neutron detection are therefore not necessarily a good choice for high-energy photon detection. At the same time, scintillators used traditionally in camera-based X-ray detectors are not necessarily a good choice either, due to the different working principles and especially the higher timing resolution achievable with LumaCam detectors. It is therefore important to re-evaluate promising neutron and traditional X-ray scintillator screens, as well as to investigate new types of scintillators for this application. In this work we present detailed results from measurements with different scintillators and gauge their potential for the use in LumaCam detectors. The aim is not to perform fundamental scintillator characterization but to evaluate known scintillators for this specific application.
    
\section{Experimental Setup}

    \begin{figure}
        \centering
        \includegraphics[width=0.95\linewidth]{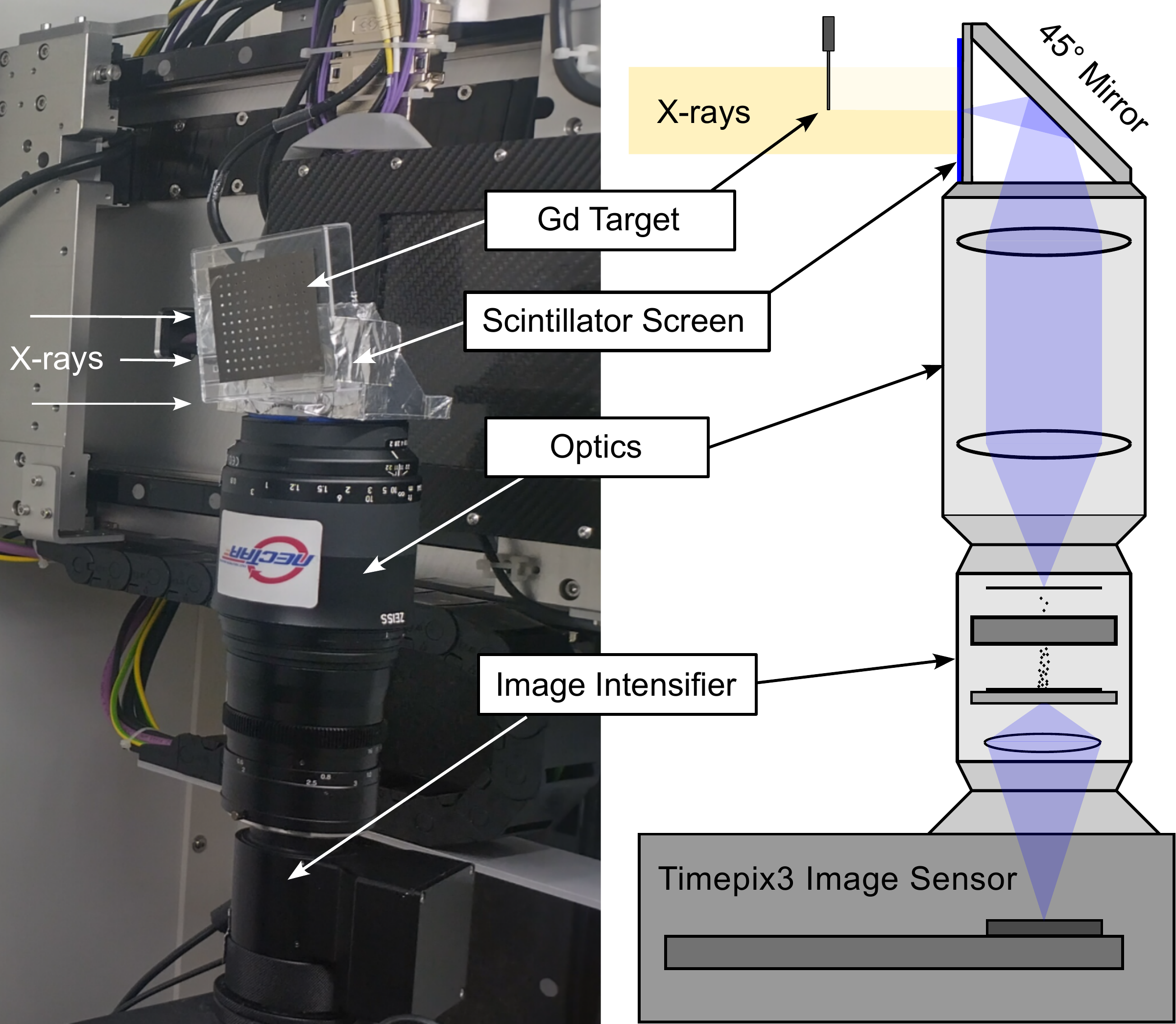}
        \caption{Image and sketch of the detector setup used to acquire the X-ray image. The X-ray source is outside the image to the left.}
        \label{fig:xrayimage_setup}
    \end{figure}

    \subsection{LumaCam Detector}
    \label{sec:setup_lumacam}

        Two LumaCam detectors were used to collect the data presented here. A picture and a schematic drawing of one of the detectors is shown in figure~\ref{fig:xrayimage_setup}. Both detectors are similar to the detector described by Wolfertz et al.~\cite{Wolfertz2024}. They have a readout chip Timepix3, which features a $256 \times 256$ pixel matrix with a pitch of $55$~{\textmu}m in both directions~\cite{poikela_timepix3_2014}. The chip is bump-bonded to a silicon optical sensor with high quantum efficiency~\cite{timepixcam, Nomerotski2019} and read out with a SPIDR system~\cite{Visser_2015}. Each pixel incorporates processing electronics that record the time of arrival (ToA) with a timing resolution of $1.56$~ns, together with the time-over-threshold (ToT), which is proportional to the energy deposited in the pixel \cite{Zhao2017}. For single-photon-sensitive operation, the camera is coupled to an image intensifier, a vacuum device equipped with a high-quantum-efficiency green-sensitive (hi-QE-green) photocathode, followed by a double (chevron) microchannel plate (MCP) and a P47 scintillator. The intensifier, its power supply, and back-side optics are integrated in a Cricket${\text{TM}2}$~\cite{photonis_cricket2}, which was mounted in front of the camera sensor.
        
        %Timepix3 chip as image sensor~\cite{medipix3} combined with a SPIDR readout board~\cite{Visser_2015}, and a Cricket2~\cite{photonis_cricket2} in the HI-QE green, double MCP, and P46 configuration from Photonis as image intensifier.

        One of the detectors is built for detailed analysis of different scintillators and uses optics with $2\times$~magnification between the scintillator and image intensifier to provide high spatial resolution, resulting in improved insight into the internal structure of individual scintillation events. It is built with a flexible holder for different scintillator screens to analyze. This configuration is called ``scintillator study configuration'' in the remainder of this work. The other detector is built to showcase the imaging application potential with a larger field-of-view. This is achieved by using optics with $2\times$~de-magnification between the scintillator and image intensifier. It uses a gadolinium oxysulfide powder doped with cesium, praseodym, and fluorine (Gd$_2$O$_2$S:Ce,Pr,F) as scintillator material. Three different scintillator screens from RC Tritec~\cite{rctritec_gdscintillators} with thicknesses of $50$~{\textmu}m, $100$~{\textmu}m, and $180$~{\textmu}m were used. This configuration is called ``imaging configuration'' in the remainder of this work.

        \begin{table*}
            \centering
            \begin{ruledtabular}
            \begin{tabular}{c c c c c c}
                \toprule
                Symbol             & Description                                                       & Scint.   & Img. \\
                \hline
                \multicolumn{4}{l}{scintillation photon reconstruction stage}\\
                \hline
                $d_{\text{px},s}$  & max. pixel activation distance in space                           & $2$~px   & $2$~px \\
                $d_{\text{px},t}$  & max. pixel activation distance in time                            & $100$~ns & $50$~ns\\
                $k$                & min. number of pixel activations per scintillation photon         & $2$      & $2$\\
                \hline
                \multicolumn{4}{l}{high-energy photon reconstruction stage}\\
                \hline
                $d_{\text{ph},s}$  & max. scintillation photon distance in space                       & var.     & $1$~px\\
                $d_{\text{ph},t}$  & max. scintillation photon distance in time                        & var.     & $5$~{\textmu}s\\
                $D_t$              & max. scintillation photon cluster duration                        & var.     & $25$~{\textmu}s\\
                $Q$                & min. high-energy photon event PSD parameter value                 & var.     & $50$~ns \\
                $m$                & min. number of scintillation photons per high-energy photon event & var.     & $2$\\
                \bottomrule
            \end{tabular}
            \end{ruledtabular}
            \caption{List of the parameter values for the event reconstruction used with the scintillator study configuration (Scint.) and the imaging configuration (Img.). The values for the high-energy photon reconstruction step for the scintillator study configuration are changed between measurements to match the individual scintillator.}
            \label{tab:parameters_reconstruction}
        \end{table*}

    \subsection{Event Reconstruction Algorithm}
    \label{sec:setup_eventReconstructionAlgorithm}
        
\begin{figure*}
            \begin{subfigure}{0.47\linewidth}
                \caption{fast (YAG:Ce) with $^{137}$Cs source}
                \includegraphics[width=\linewidth]{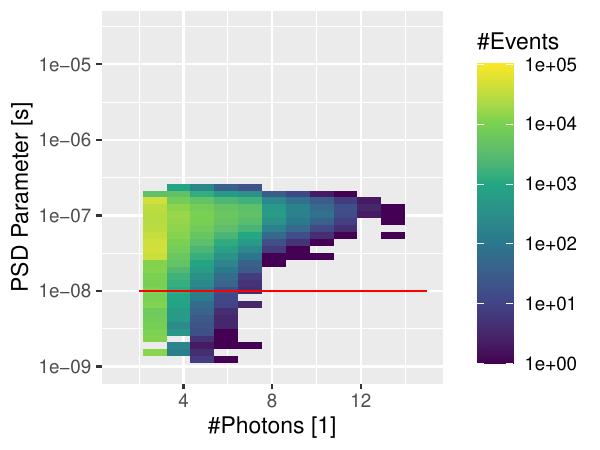}
            \end{subfigure}
            \begin{subfigure}{0.47\linewidth}
                \caption{slow (Gd$_2$O$_2$S:Ce,Pr,F) with $^{137}$Cs source}
                \includegraphics[width=\linewidth]{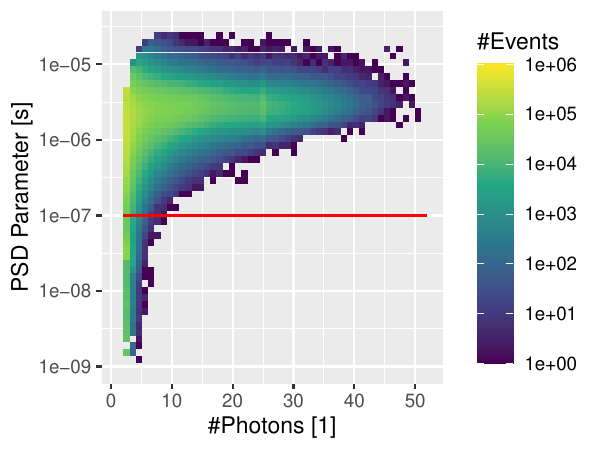}
            \end{subfigure}\\
            \begin{subfigure}{0.47\linewidth}
                \caption{fast (YAG:Ce) without source}
                \includegraphics[width=\linewidth]{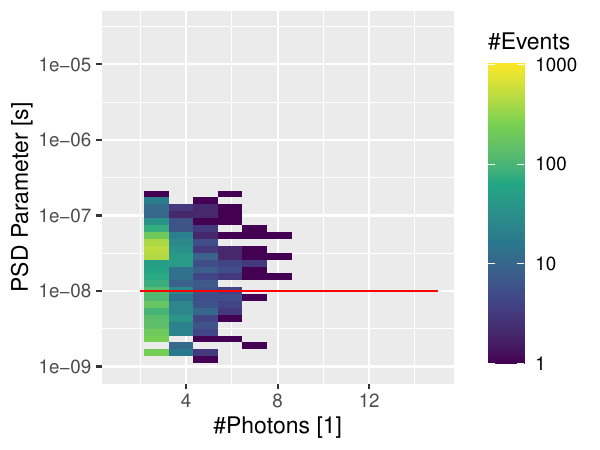}
            \end{subfigure}
            \begin{subfigure}{0.47\linewidth}
                \caption{slow (Gd$_2$O$_2$S:Ce,Pr,F) without source}
                \includegraphics[width=\linewidth]{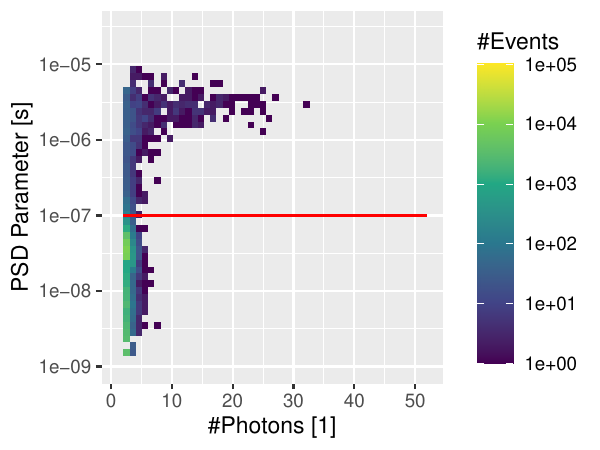}
            \end{subfigure}
            \caption{Example of how the PSD parameter threshold acts demonstrated on a faster (YAG:Ce in (a) and (c)) and a slower (Gd$_2$O$_2$S:Ce,Pr,F in (b) and (d)) scintillator exposed to a $^{137}$Cs source ((a) and (b)) and without a source present ((c) and (d)). The shape of the region below the threshold (red line) does not change significantly between the measurement with and without the source, indicating a large noise contribution for events falling in this region.}
            \label{fig:yagce_nPhotonsVsPsd}
        \end{figure*}

        The algorithm for the reconstruction of individual high-energy photon events is based on the dual-stage clustering algorithm described in by Wolfertz et~al.~\cite{Wolfertz2024}. During the first stage, the scintillation photons activating the image intensifier are reconstructed from the hit pixels on the Timepix3 imaging chip by first clustering these pixel hits and then filtering the clusters by the number of pixel activations in them. During the second stage, the high-energy photons (or other particles such as neutrons) are reconstructed from the scintillation photons by a second clustering and filtering step. The only adaptation made to the reconstruction algorithm for the work presented here is an additional pulse-shape discrimination (PSD) criterion for the filter of the high-energy photon event reconstruction stage. A PSD parameter is calculated for each cluster of scintillation photons. The PSD parameter $q_A$ for a set of photons $A$ belonging to a single cluster is defined as the average distance in time of all photons to the first photon in this cluster:
        \begin{equation}
            q_A = \frac{1}{N_A}\sum_{p \in A}t_p - t_0(A)
        \end{equation}
        Where $N_A = \vert A \vert$ is the number of photons in $A$, $t_p$ is the arrival time of a photon $p$, and $t_0(A) = \text{min}\left( \{t_p \mid p \in A \} \right)$ is the arrival time of the first photon in $A$. A scintillation photon cluster is only accepted to correspond to a high-energy photon event if the PSD parameter for that cluster is above a threshold $Q$. Figure~\ref{fig:yagce_nPhotonsVsPsd} demonstrates how this can work to remove noise from the final detector signal. The complete list of parameters for the high-energy photon event reconstruction algorithm are summarized in table~\ref{tab:parameters_reconstruction}.

        %The parameters used with the scintillator study configuration for the scintillation photon reconstruction step of the algorithm were: $d_{\text{px},s} = 2$~px and $d_{\text{px},t} = 100$~ns for the maximum pixel activation distance in space and time respectively, and $k = 2$ for the minimum number of pixel activations per photon. The parameters for the high-energy photon event reconstruction step were varied to match the different scintillators. The parameters used with the imaging configuration were: $d_{\text{px},s} = 2$~px and $d_{\text{px},t} = 50$~ns for the maximum pixel activation distance in space and time respectively, $k = 2$ for the minimum number of pixel activations per scintillation photon, $d_{\text{ph},s} = 1$~px for the maximum scintillation photon distance in space, $d_{\text{ph},t} = 5$~{\textmu}s for the maximum scintillation photon distance in time, $D_t = 25$~{\textmu}s for the maximum event duration, $Q = 50$~ns for the minimum PSD parameter, and $m = 2$ for the minimum number of scintillation photons per high-energy photon event. \textcolor{red}{- this needs better explanations, i am struggling to understand what is $d_{\text{px},s}$ and $d_{\text{px},t}$ - }
        
        %\begin{table}[]
        %    \centering
        %    \begin{tabular}{c l}
        %        \toprule
        %        Symbol / Definition        & Description                     \\
        %        \midrule
        %        $x_{a(A)}$                 & $x$ coordinate of the position \\
        %        $y_{a(A)}$                 & $y$ coordinate of the position \\
        %        $t_{a(A)}$                 & time                           \\
        %        $n_{a(A)} = \vert A \vert$ & number of photons              \\
        %        $q_{a(A)} = q_A$           & PSD parameter                  \\
        %        \bottomrule
        %    \end{tabular}
        %    \caption{List of the properties for each reconstructed scintillation event $a(A)$, where $A$ is the corresponding set of photons}
        %    \label{tab:parameters_particleEvent}
        %\end{table}

        %\begin{itemize}
        %    \item $x_{a(A)}$ : $x$ coordinate of the position
        %    \item $y_{a(A)}$ : $y$ coordinate of the position
        %    \item $t_{a(A)}$ : time
        %    \item $n_{a(A)} = \vert A \vert$ : number of photons
        %    \item $q_{a(A)} = q_A$ : PSD parameter
        %\end{itemize}

        %\begin{itemize}
        %    \item $d_{\text{px},s}$ : maximum pixel activation distance space
        %    \item $d_{\text{px},t}$ : maximum pixel activation distance time
        %    \item $k$ : minimum number of pixel activations
        %    \item $d_{\text{ph},s}$ : maximum photon distance space
        %    \item $d_{\text{ph},t}$ : maximum photon distance time
        %    \item $D_t$ : maximum event duration
        %    \item $m$ : minimum number of photons
        %    \item $Q$ : minimum PSD parameter
        %\end{itemize}
        
    \subsection{X-Ray Image}
    \label{sec:setup_xrayimage}

        To acquire test images with X-rays, the detector was installed in the imaging configuration (see section~\ref{sec:setup_lumacam}) in a Tescan UniTom HR X-ray imaging device~\cite{tescan_uniTom}, directly in front of the device's integrated detector. The LumaCam detector was mounted with the scintillator screen directly in front of the sensitive area of the built-in detector of the X-ray device. A $100$~{\textmu}m thick sheet of gadolinium with $1$~mm square holes was placed in front of the detector as a test sample. Figure~\ref{fig:xrayimage_setup} shows a picture of the setup. The Tescan UniTom HR X-ray imaging device typically employs so-called cone-beam magnification, where the projection of the sample is larger than the sample itself due to the source being smaller than the sample. It can be used to improve the resolution  of a radiograph beyond the detector's limit. The geometry chosen here almost completely eliminates any cone beam magnification, as the magnification would also make any imperfections of the sample more pronounced and would therefore make it harder to accurately determine the detector resolution.

        \begin{table}
            \centering
            \begin{ruledtabular}
            \begin{tabular}{c c c c}
                Scintillator    & Source  & \multicolumn{2}{c}{Time}\\
                                            \cmidrule(lr){3-4}
                Thickness       & Power   & With Sample  & Open Beam \\
                \hline
                 $50$~{\textmu}m & $6.5$~W & $900$~s & $1800$~s  \\
                $100$~{\textmu}m & $4.5$~W & $600$~s &  $600$~s  \\
                $180$~{\textmu}m & $3.5$~W & $300$~s &  $300$~s  \\
            \end{tabular}
            \end{ruledtabular}
            \caption{Acquisition parameters for the different X-ray image measurements.}
            \label{tab:xrayimage_acqParameters}
        \end{table}

        The detector was configured with each of the three Gd$_2$O$_2$S:Ce,Pr,F scintillator screens using them one at a time. The reconstructed X-ray events from the detector were binned into a $512 \times 512$ grid based on their spatial position to create an image. With each scintillator screen, an image with (called sample image) and without (called open beam image) the test sample was recorded. The source current was adjusted for each setup to prevent detector saturation with the thicker, more efficient scintillator screens. The exact power and acquisition times for each measurement are shown in table~\ref{tab:xrayimage_acqParameters}. The acceleration voltage for the X-ray source was set to $120$~kV for all measurements.

    \subsection{Scintillator Study}
    \label{sec:setup_scintillatorStudy}

        In this study, we employed four different scintillator materials: \textbf{Gd$_2$O$_2$S:Ce,Pr,F} (Gadolinium oxysulfide doped with cerium, praseodymium, and fluorine, fabricated by RC Tritec~\cite{rctritec_gdscintillators}), \textbf{YAP:Ce} (Yttrium Aluminum Perovskite doped with Cerium), \textbf{GAGG+} (Gadolinium Aluminum Gallium Garnet), and \textbf{YAG:Ce} (Yttrium Aluminum Garnet doped with Cerium).

        \subsubsection{Properties of the Scintillators}

            The selection of these four scintillators was based on their distinct  physical and scintillation properties, which are presented below and summarized in table~\ref{tab:scintillatorProperties}.
    
            \begin{itemize}
                \item \textbf{Gd$_2$O$_2$S:Ce,Pr,F} was selected as an established powder scintillator, previously used with LumaCam detectors\cite{MIRASHI2025170284}, to serve as a comparison.
                \item \textbf{YAP:Ce} was selected for its \textit{moderate light yield and fast response time}, making it suitable for applications requiring high temporal resolution.
                \item \textbf{GAGG+} was chosen for its \textit{very high light yield (about 45,000 photons/MeV) and good stopping power}, making it effective for gamma-ray detection.
                \item \textbf{YAG:Ce} was included due to its \textit{high mechanical stability and broad optical compatibility}, making it a robust option. This scintillator is used in optical intensifiers (in powder form) and is known as P46.
            \end{itemize}

            \begin{table*}
                \centering
                \begin{ruledtabular}
                \begin{tabular}{c S c c c c c}
                    Scintillator & {Density $\left[\frac{\text{g}}{\text{cm}^3}\right]$} & Light Yield $\left[\frac{\text{photons}}{\text{MeV}}\right]$ & Decay Time [ns] & Peak Emission [nm] & Thickness [{\textmu}m]\\
                    \hline
                    Gd$_2$O$_2$S:Ce,Pr,F & 7.34* & 35000 & 4000 & 510 & 180\\
                    %Gd$_2$O$_2$S:Tb     & 7.34* &  ???  & 3000 & 550 \
                    YAP:Ce               & 5.37  & 25000 &   25 & 370 & 400\\
                    GAGG+                & 6.7   & 45000 &  120 & 520 & 50\\
                    YAG:Ce               & 4.57  & 30000 &   70 & 550 & 100\\
                \end{tabular}
                \end{ruledtabular}
                \caption{Properties of the scintillator screens used in the scintillator study. The density of Gd$_2$O$_2$S:Ce,Pr,F is the density of the pure scintillator material, excluding any binder material to hold the powder together.~\cite{Carel_W_E_van_Eijk_2002, crytur}}
                \label{tab:scintillatorProperties}
            \end{table*}

\begin{figure}
                \centering
                \includegraphics[width=0.8\linewidth]{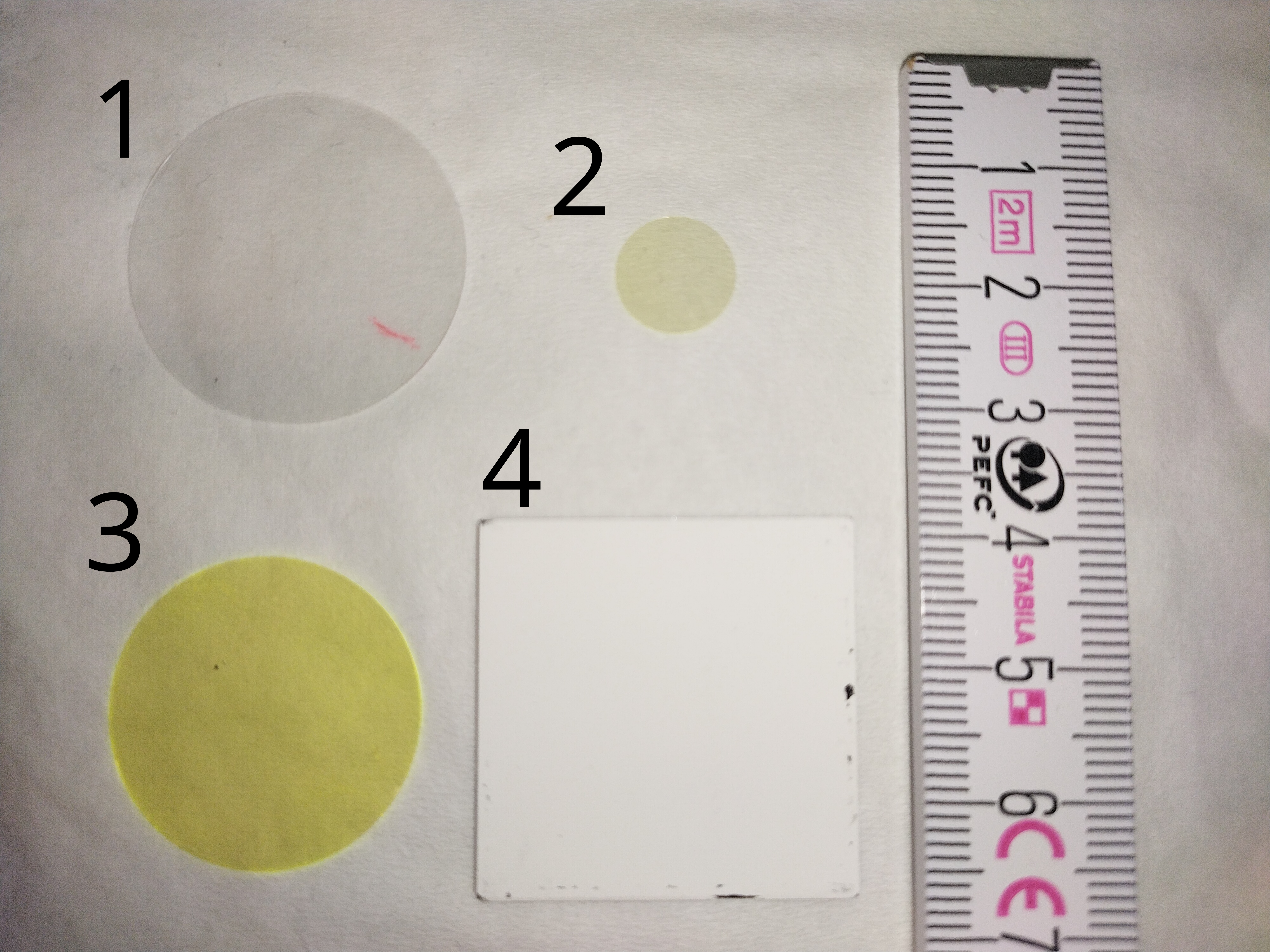}
                \caption{Photograph of the scintillators used for the scintillator study. 1 is YAP:Ce, 2 is GAGG+, 3 is YAG:Ce, and 4 is Gd$_2$O$_2$S:Ce,Pr,F. The numbers on the ruler are in cm.}
                \label{fig:sinctillators_picture}
            \end{figure}

            The single-crystal scintillators used in this study were grown by Crytur \cite{crytur} using the Czochralski method\cite{czochralskiNeuesVerfahrenZur1918}. This technique involves melting the raw materials in a crucible and slowly pulling a seed crystal from the melt while rotating it. This controlled process ensures high crystal purity, uniformity, and minimal defects, which are crucial for achieving optimal scintillation performance. The resulting single crystals were precisely cut and polished.  Figure~\ref{fig:sinctillators_picture} shows a photograph of the scintillator screens and the thicknesses of each screen is specified in table~\ref{tab:scintillatorProperties}.

            Other than YAP:Ce, all scintillators have a high radiation hardness. Their combined use allowed for a comprehensive evaluation of X-ray and gamma-ray detection \cite{niklScintillationDetectorsXrays2006} with the LumaCam optical camera across different energy ranges.
    
            %\begin{table}[h]
            %    \centering
            %    \caption{Properties of scintillators used in the study}
            %    \label{tab:scintillators}
            %    \begin{tabular}{|l|c|c|c|c|c|}
            %        \hline
            %        \textbf{Property} & \textbf{YAP:Ce} & \textbf{GAGG+} & \textbf{YAG:Ce} & \textbf{LuAG:Pr} & \textbf{Gd2O2S}\\
            %        \hline
            %        \textbf{Density (g/cm³)} & 5.37 & 6.7 & 4.57 & 6.73 & 7.34 (pure material)\\
            %        \textbf{Light Yield (photons/MeV)} & 25,000 & 45,000 & 30,000 & 18,000 & xxxxx\\
            %        \textbf{Decay Time (ns)} & 25 & 120 & 70 & 20 & 3000\\
            %        \textbf{Peak Emission Wavelength (nm)} & 370 & 520 & 550 & 310 & 550 \\
            %        \textbf{Radiation Hardness} & Limited & High & High & High & ??\\
            %        \textbf{Hygroscopicity} & No & No & No & No & No\\
            %        \textbf{Mechanical Stability} & High & High & High & High & High\\
            %        \hline
            %    \end{tabular}
            %\end{table}
            
\subsubsection{Measurements with Gamma Sources}

    To investigate the detector response for different scintillator materials, measurements were performed using the scintillator study configuration described in  section~\ref{sec:setup_lumacam}, with the scintillator screens installed in the detector one at a time. Three measurements were performed with each scintillator screen: one with a $^{137}$Cs source (gamma energy of $662$~keV), one with a $^{241}$Am source (main gamma energy of $59.5$~keV), and one without any source. The sources were placed directly in front of the scintillator each time. The parameters for the high-energy photon event reconstruction step (see section~\ref{sec:setup_eventReconstructionAlgorithm}) were varied to match the different scintillator properties, and are listed in table~\ref{tab:parameters_reconstruction_scintillatorStudy}.

    \begin{table}
        \centering
        \begin{ruledtabular}
        \begin{tabular}{c c c c c}
            Symbol             & Gd$_2$O$_2$S:Ce,Pr,F & YAP:Ce           & GAGG+            & YAG:Ce           \\
            \hline
            $d_{\text{ph},s}$  & $40$~px              & $40$~px          & $40$~px          & $40$~px          \\
            $d_{\text{ph},t}$  & $50$~{\textmu}s      & $10$~{\textmu}s  & $5$~{\textmu}s   & $5$~{\textmu}s   \\
            $D_t$              & $5$~ms               & $100$~{\textmu}s & $500$~{\textmu}s & $500$~{\textmu}s \\
            $Q$                & $100$~ns             & 0                & $10$~ns          & $10$~ns          \\
        \end{tabular}
        \end{ruledtabular}
        \caption{List of the parameter values for the high-energy photon event reconstruction step for the scintillator study}
        \label{tab:parameters_reconstruction_scintillatorStudy}
    \end{table}

\section{Results and Discussion}

    \subsection{X-Ray Imaging}

\begin{figure*}
            \centering
            \begin{subfigure}{0.49\linewidth}
                \caption{}
                \includegraphics[width=\linewidth]{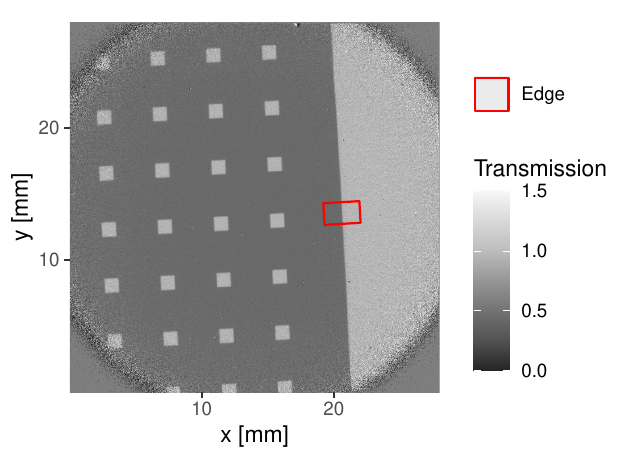}
                \label{fig:xrayimage}
            \end{subfigure}
            \begin{subfigure}{0.4\linewidth}
                \caption{}
                \includegraphics[width=\linewidth]{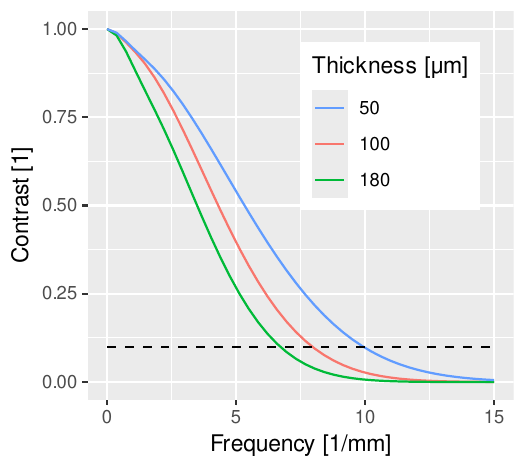}
                \label{fig:xrayimage_mtf}
            \end{subfigure}
            \caption{(a): X-ray transmission image of the test sample. The region used to fit the edge spread function is marked in red; (b): Modulation transfer function measured for the different scintillator thicknesses. The $0.1$ contrast level used to calculate the resolution is marked with a black dashed line.}
        \end{figure*}

        The X-ray images collected with the setup described in section~\ref{sec:setup_xrayimage} are normalized to the acquisition time. The normalized sample images are divided by the corresponding normalized open beam images to obtain a test sample transmission image for each scintillator thickness. Figure~\ref{fig:xrayimage} shows the transmission image for the $50$~{\textmu}m thick scintillator screen. Due to the mismatch of the image intensifier geometry (a circle with a diameter of $18$~mm), and the image sensor (a square with a edge length of $14$~mm), the corners of the image sensor are not illuminated. Apart from the corners, the image normalizes well and shows a good overall quality.

        The detector resolution is estimated via the edge spread function measured at the outer edge of the test sample. The transmission profile perpendicular to the edge is fitted to the sum of two gauss error functions. The modulation transfer function (MTF) is obtained by taking the absolute value of the Fourier transformation of the derivative of the fit function and normalizing the result to be 1 at its peak. The MTF for each scintillator screen is shown in figure~\ref{fig:xrayimage_mtf}. The resolution is calculated by taking half of the period at which the MTF reaches $0.1$. The resulting estimated resolution is $50$~{\textmu}m for the $50$~{\textmu}m thick scintillator, $63$~{\textmu}m for the $100$~{\textmu}m thick scintillator, and $74$~{\textmu}m for the $180$~{\textmu}m thick scintillator.

        As is expected, the scintillator screen thickness has a significant influence on the achievable resolution. This is likely predominantly due to the spread of scintillation light in the diffusely scattering scintillator material. It is important to note that the measured resolution is not a direct quantitative indicator of the scintillation light spread. The event reconstruction algorithm (see section~\ref{sec:setup_eventReconstructionAlgorithm}) is generally capable of providing a better accuracy in the reconstructed position than the spread of the photons. At the same time, the size of the physical pixels on the sensor can limit the achievable resolution. In this configuration, each pixel corresponds to a $110$~{\textmu}m$\times 110$~{\textmu}m area on the scintillator screen. In the case of the $50$~{\textmu}m thick screen, the detector has a resolution better than half a pixel size, demonstrating the super-resolution capabilities of LumaCam detectors. In case of the thicker scintillator screens, the resolution was better than the scintillator thickness, demonstrating how the event-based reconstruction compensates the effect of light diffusion.

        A scintillator screen thinner than $50$~{\textmu}m would likely provide even better resolution, but at the cost of lower X-ray interaction probability. However, it is expected, that at some point, the size of the physical sensor pixels will be a limiting factor. Significant resolution improvements could still be made past this point by switching to higher magnification optics, at the cost of a smaller field-of-view.

    \subsection{Comparison of Scintillators}

        \subsubsection{Scintillation Photon Multiplicity}
    
\begin{figure*}
                \begin{subfigure}{0.4\linewidth}
                    \caption{Gd$_2$O$_2$S:Ce,Pr,F}
                    \includegraphics[width=\linewidth]{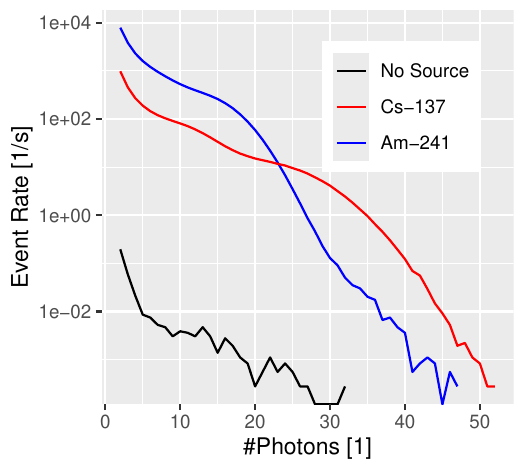}
                \end{subfigure}
                \begin{subfigure}{0.4\linewidth}
                    \caption{YAP:Ce}
                    \includegraphics[width=\linewidth]{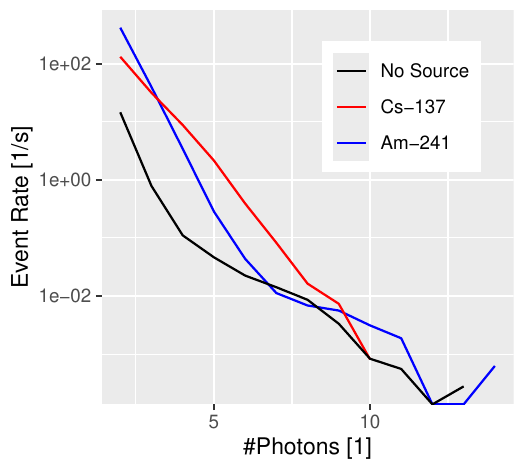}
                \end{subfigure}
                \begin{subfigure}{0.4\linewidth}
                    \caption{GAGG+}
                    \includegraphics[width=\linewidth]{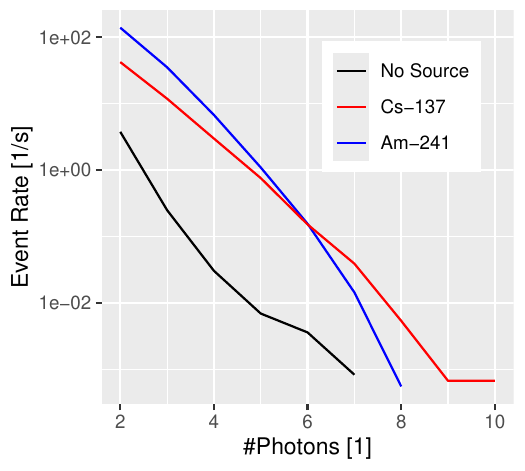}
                \end{subfigure}
                \begin{subfigure}{0.4\linewidth}
                    \caption{YAG:Ce}
                    \includegraphics[width=\linewidth]{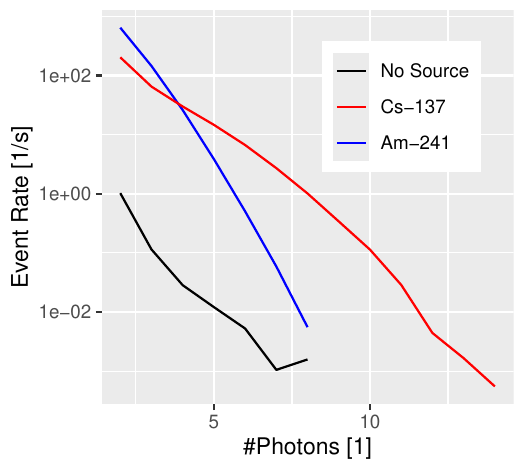}
                \end{subfigure}
                \caption{Photon multiplicity spectra of the different scintillators exposed to different radioactive sources. The data for Gd$_2$O$_2$S:Ce,Pr,F shows a significantly higher measured photon multiplicity than the others due to its comparatively long decay time. All scintillators except YAP:Ce show a significant gap between the rates recorded with either source and without a source present over the entire recorded spectrum. Gd$_2$O$_2$S:Ce,Pr,F and YAP:Ce show a significant shape difference in their spectrum between the two sources, indicating potential for energy-sensitive measurements. Due to the uncertainties in the measurement setup, a detailed quantitative comparison of the absolute rates is not possible.}
                \label{fig:photonSpectra}
            \end{figure*}

            An important aspect to consider for scintillator candidates is the number of scintillation photons detected per high-energy photon event (scintillation photon multiplicity). Events with fewer scintillation photons generally result in less accurate values for the reconstructed properties of the high-energy photon. Setting a good photon threshold value is important for the noise rejection effectiveness. In addition, the scintillation photon multiplicity is dependent on the energy deposited into the scintillator screen, and knowledge on this relation could enable gathering information on the energy spectrum of the incoming X-rays or gamma rays based on the measured scintillation photon multiplicity spectrum. If the scintillation photon flux is too high, it can saturate the detector. Especially in high flux applications, a larger number of scintillation photons per event does not provide any benefit as the light from the scintillator screen has to be artificially attenuated more to prevent detector saturation. Figure~\ref{fig:photonSpectra} shows the scintillation photon multiplicity spectra recorded for the four types of scintillators and with the two types of radioactive sources as well as without a source present.

            The Gd$_2$O$_2$S:Ce,Pr,F screen produces by far the most scintillation photons per high-energy photon event. This is expected as it has a slower decay time than the other scintillators, which reduces the chance of several photons in a single event being too close together for the detector to separate and therefore being counted as a single photon. The events in the Gd$_2$O$_2$S:Ce,Pr,F scintillator generally have a high enough scintillation photon multiplicity that the scintillator will also work well with less light-efficient optics. For applications with a high X-ray flux, it would even be desirable to use lower light efficiency optics to prevent detector saturation (the measurements described in section~\ref{sec:setup_xrayimage} for example use optical attenuators to purposefully lower the efficiency of the optics and thereby allow measuring at higher X-ray detection rates).
            
            %The count rate for the Gd$_2$O$_2$S:Ce,Pr,F rises steeply below ${\sim}5$ photons, which might indicate that a significant portion of events below $5$ photons are not real events from gamma rays but some kind of noise associated with real events, e.g. connected to the afterglow tail. A threshold $m$ of ${\sim}5$ photons would have to be set to exclude these events.
            
            The events for the other three scintillators show a significantly lower measured scintillation photon multiplicity than the ones for Gd$_2$O$_2$S:Ce,Pr,F. The difference is significantly higher than the difference in light yield between the scintillators (see table~\ref{tab:scintillatorProperties}). Apart from the influece from the difference in the decay time described above, the different ways in which light transport happens in powder and single crystal scintillator screens could also be a reason for this apparent discrepancy. Due to the uncertainty in the gamma flux during these measurements (the source being very close to the scintillator screen), a detailed quantitative comparison between the scintillators is not possible. All of the single crystal scintillators should be used with high efficiency optics, such as the one used in this measurements, and they might even profit from higher efficiency optics.

            The count rate for Gd$_2$O$_2$S:Ce,Pr,F with the sources is many orders of magnitude above the background without a source present, indicating that the background noise rejection is working well and almost all of the signal comes from the gamma rays. The signal with the sources for GAGG+ and YAG:Ce is also clearly separated from the background. While the difference is not as strong as for Gd$_2$O$_2$S:Ce,Pr,F, this still indicates that the background noise rejection is working well and most of the signal comes from the gamma rays. 
            
            The separation between the rates with and without a source is less strong for YAP:Ce, especially for the lower energy $^{241}$Am gamma rays, indicating that a non-negligible portion of the signal with the sources present is actually due to background noise. Another contributing factor to the lower signal yield for this scintillator could be a mismatch in the photocathode efficiency of the image intensifier since its peak emission is at a lower wavelength compared to the other scintillators, see table \ref{tab:scintillatorProperties}.

            There is a significant difference in the shape of the scintillation photon multiplicity rates for Gd$_2$O$_2$S:Ce,Pr,F between the two sources with a relative change of ${\sim}300\times$ between $15$ photons ($^{241}$Am ${\sim}10\times$ higher than $^{137}$Cs) and $30$ photons ($^{137}$Cs ${\sim}30\times$ higher than $^{241}$Am). This shows a significant dependence on the gamma energy and indicates that it can be a potentially useful scintillator for energy discrimination applications. YAG:Ce also exibits a significant difference in the shape between the two sources with a factor of ${\sim}300\times$ between $3$ photons ($^{241}$Am ${\sim}3\times$ higher than $^{137}$Cs) and 8 photons ($^{137}$Cs ${\sim}100\times$ higher than $^{241}$Am). It is therefore also a promising candidate to investigate further for energy discrimination applications. The shapes for the two different sources for YAP:Ce and GAGG+ are much closer together. They are therefore unlikely to be good candidates for energy discrimination applications.

        \subsubsection{Temporal Resolution}

            In the tested detection scheme, an incident X-ray generates a burst of optical photons in the scintillator, which are individually time-stamped by the LumaCam event reconstruction algorithm. The photon emission times follow the intrinsic scintillation pulse shape, characterized by rise and decay constants, and are convolved with the timing response of the image intensifier and readout electronics.
            
            The event time $t$ of a high-energy photon event reconstructed from a set of scintillation photons $A$ is defined as the arrival time of the first detected scintillation photon:
            \begin{equation}
                t = t_0(A) = \min \{ t_p \, | \, p \in A \}.
            \end{equation}
            The temporal resolution is therefore determined by the statistical fluctuations of this first photon time. For an exponential scintillation decay with time constant $\tau$, the standard deviation $\sigma_t$ of the first photon time scales approximately as
            \begin{equation}
                \sigma_t(N_A) \propto \frac{\tau}{N_A},
            \end{equation}
            where $N_A = \vert A \vert$ is the number of reconstructed photons in the event. Thus, improved timing is achieved for larger photon multiplicities and shorter decay times.
            
            Assuming a purely exponential decay, the decay constant $\tau$ is equal to the expectation value of the average distance between the first photon in an event and all subsequent ones:
            \begin{equation}
                \label{eq:decayConstant_exp}
                \tau = \text{E}\left[\frac{1}{N_A - 1}\sum_{p \in A} t_p - t_0(A)\right]
            \end{equation}
            Here, $\text{E}\left[X\right]$ denotes the expected value of $X$. This property allows a straightforward experimental estimation of the effective decay time from the data. However, it is important to note that in practice, the decay is usually not purely exponential, and frequently not all photons belonging to an event are correctly associated with this event. The measured decay constant is therefore only an approximation. 
            
\begin{figure}
                \centering
                \begin{subfigure}{0.9\linewidth}
                    \caption{}
                    \includegraphics[width=\linewidth]{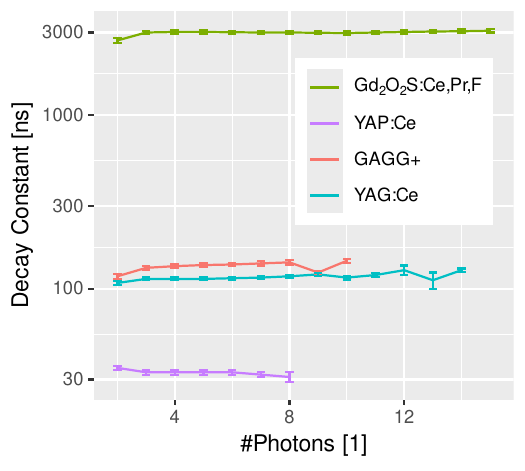}
                    \label{fig:decayConstant_measured}
                \end{subfigure}
                \begin{subfigure}{0.9\linewidth}
                    \caption{}
                    \includegraphics[width=\linewidth]{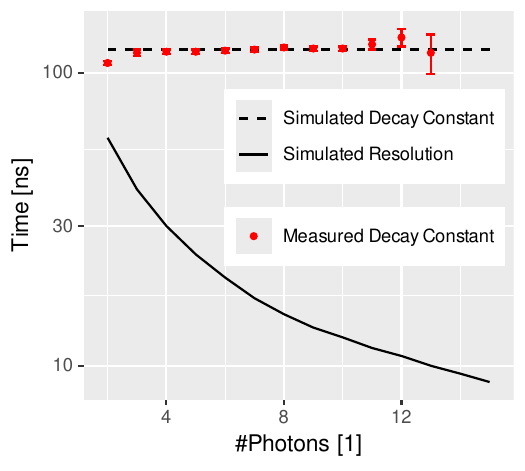}
                \end{subfigure}
                \caption{(a): Measured decay constant as a function of the number of detected photons for the investigated scintillators illuminated with 662 keV gamma rays from a $^{137}$Cs radioactive source; (b): comparison of simulated and measured decay constants and simulated prediction of time resolution, all for the YAG:Ce scintillator.}
                \label{fig:scintillatorPSD}
            \end{figure}

            Figure~\ref{fig:decayConstant_measured} shows the measured decay constant as a function of the number of detected photons $N_A$ for the investigated scintillators illuminated with 662 keV gamma rays from a $^{137}$Cs radioactive source. The values are approximately constant with respect to $N_A$, indicating that the effect of the the imperfect photon association to the events is relatively small. From these measurements we extract effective decay times of 34~ns (YAP:Ce), 118~ns (YAG:Ce), 139~ns (GAGG+), and 3076~ns (Gd$_2$O$_2$S:Ce,Pr,F). These values are, in general, consistent with the known scintillation properties of the materials as presented in table \ref{tab:scintillatorProperties} with deviations, which can be attributed to differences in the exact composition of the used scintillators.
            
            To relate photon multiplicity and properties of scintillators to timing performance, we implemented a simple Monte Carlo simulation. Photon emission times are sampled from an exponential distribution with decay constant $\tau$. For each event, $N$ photons are randomly selected, and both the average distance from the first photon to the subsequent ones and the delay between the known beginning of the simulated event and the first photon are calculated. For the purposes of this study we neglected timing effects due to the image intensifier and temporal transfer function of the Timepix3 front-end electronics, as their contributions would be important only for the sub-nanosecond scale timing \cite{Nomerotski2023, Heijhoff2022, Ballabriga2023}.
            
            The results are shown in figure~\ref{fig:scintillatorPSD}b) for the YAG:Ce scintillator. The value of 118~ns (as measured in the data) was assumed for the decay time $\tau$ of the simulated sample. The simulated timing resolution, defined as the standard deviation of the first photon time, follows the expected $1/N$ dependence. We also checked that a small rise time in the scintillator response, up to 10~ns, does not significantly affect the timing results if the number of detected photons is small, less than 100, as in our case here. 
            
\begin{figure}
                \centering
                \begin{subfigure}{0.9\linewidth}
                    \caption{YAG:Ce}
                    \includegraphics[width=\linewidth]{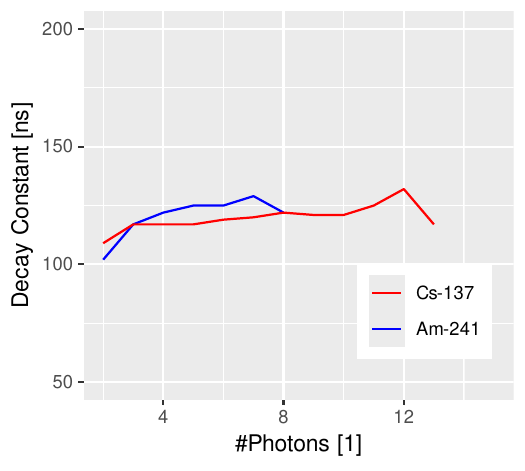}
                \end{subfigure}
                \begin{subfigure}{0.9\linewidth}
                    \caption{Gd$_2$O$_2$S:Ce,Pr,F}
                    \includegraphics[width=\linewidth]{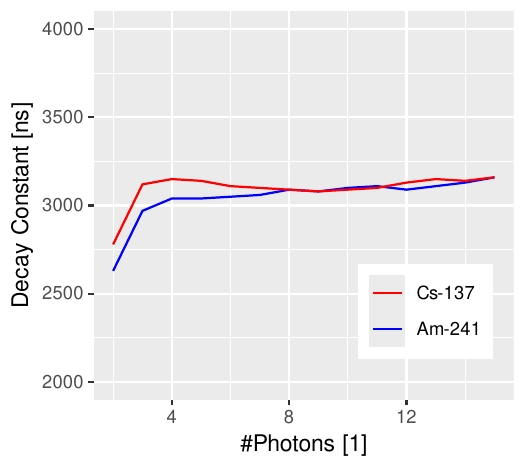}
                \end{subfigure}
                \caption{(a): Measured decay constant as a function of the number of detected photons for the YAG:Ce scintillator for $^{137}$Cs and $^{241}$Am radioactive source; (b): same for the Gd$_{2}$O$_{2}$S:Ce scintillator.}
                \label{fig:differentSources}
            \end{figure}

            We show the measured decay constant for two types of scintillators, YAG:Ce and Gd$_2$O$_2$S:Ce,Pr,F, in figure~\ref{fig:differentSources}, comparing irradiation with two different radioactive sources, $^{241}$Am and $^{137}$Cs, emitting, respectively, 59.5 keV and 662 keV gamma rays. The difference between two datasets and between different photon multiplicities can be used to estimate the systematic uncertainties of this approach. When excluding the values for $N = 2$, the maximum difference between any two measured decay constants is $12$\% for YAG:Ce and $6$\% for Gd$_2$O$_2$S:Ce,Pr,F. The drop in the measured decay constant at $N = 2$ is likely an artifact from the event reconstruction clustering algorithm. If fewer scintillation photons originating from a single high-energy photon event are detected, it becomes harder for the clustering algorithm to correctly associate late photons with the initial photons. As a result, late photons are more likely to be missed, reducing the average distance to the first photon and thereby also the measured decay constant.
            
\begin{figure}
                \centering
                \begin{subfigure}{0.9\linewidth}
                    \caption{}
                    \label{fig:nPhotons_timeResolution}
                    \includegraphics[width=\linewidth]{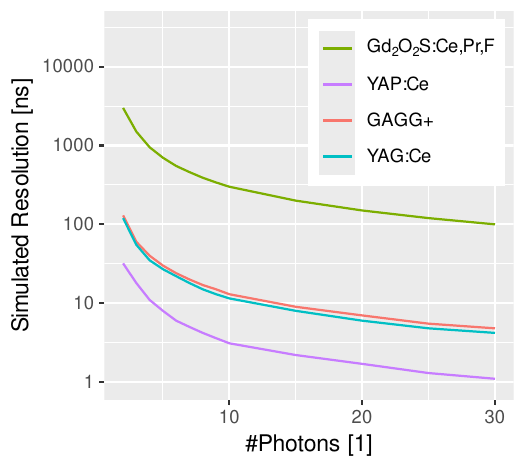}
                \end{subfigure}
                \begin{subfigure}{0.9\linewidth}
                    \caption{}
                    \label{fig:nPhotons_spectrum}
                    \includegraphics[width=\linewidth]{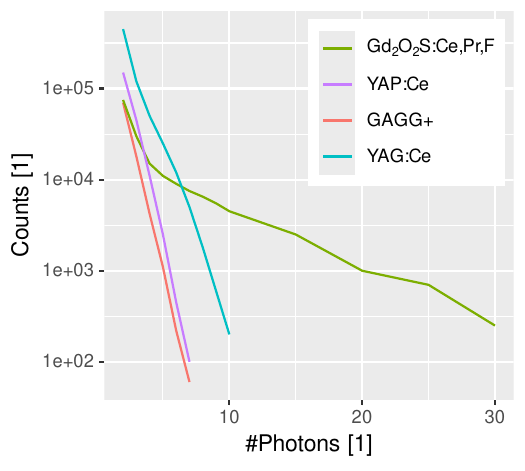}
                \end{subfigure}
                \caption{(a): Simulated time resolution as a function of photon number for different types of scintillators, using the measured decay times; (b): Distributions for the detected number of photons with $^{137}$Cs radioactive source for different types of scintillators.}
                \label{fig:nphotons}
            \end{figure}

            The overall timing performance depends on the distribution of detected photon numbers. Figure~\ref{fig:nPhotons_timeResolution} shows the simulated time resolution as a function of photon number for each scintillator, using the measured decay times. Faster scintillators provide substantially improved timing at fixed $N$, while slow scintillators remain intrinsically limited by their large decay constants.
            
            Figure~\ref{fig:nPhotons_spectrum} shows the measured photon multiplicity spectra for the four scintillators. Faster materials produce narrower distributions centered at low $N$, whereas higher light-yield or slower scintillators (much slower than the camera timing resolution of few ns) exhibit broader multiplicity spectra since the photons could be spread both in space and time so the blending effects of photon clusters will be much alleviated.
            
            To obtain a realistic estimate of timing performance, we combine simulated $N$-dependent resolution with the measured multiplicity distributions of photons calculating the weighted average time resolution with results summarized in table~\ref{tab:resolution}. The average time resolutions are 15.4~ns (YAP:Ce), 50.7~ns (YAG:Ce), 63.5~ns (GAGG+), and 941~ns (Gd$_2$O$_2$S:Ce,Pr,F). The effective photon number, defined as the ratio of decay time and predicted resolution, is approximately 2--3 for all investigated materials. Table \ref{tab:resolution} summarizes the measured decay time, time resolution and effective number of photons for four types of used scintillator. 
            
            \begin{table*}
                \centering
                \begin{ruledtabular}
                \begin{tabular}{c c c c}
                    Scintillator & Measured decay time $\tau$ (ns) & Time resolution (ns) & $N_{\mathrm{eff}}$ \\
                    \hline
                    YAP:Ce & 34   & 15.4  & 2.21 \\
                    YAG:Ce & 118  & 50.7  & 2.33 \\
                    GAGG   & 139  & 63.5  & 2.20 \\
                    Gd$_2$O$_2$S:Ce,Pr,F & 3076 & 941 & 3.26 \\
                \end{tabular}
                \end{ruledtabular}
                \caption{Measured scintillator decay time, time resolution (standard deviation), and effective photon number for different scintillators. The resolution was estimated as weighted average with the number of photons distribution taken from $^{137}$Cs source data. The effective photon number $N_{\mathrm{eff}}$ is defined as the decay time divided by the resolution.}
                \label{tab:resolution}
            \end{table*}
            
            In summary, the temporal resolution of X-ray detection with LumaCam is governed by the scintillator decay time, the number of detected photons per event, and the photon multiplicity distribution. 
            %These results demonstrate that event-by-event photon counting combined with nanosecond time stamping is able to evaluate the timing performance directly from measured photon statistics.

\section{Conclusion and Outlook}

    This work demonstrates that the LumaCam detector, originally developed for neutron imaging, is also a promising platform for event-based X-ray and gamma-ray detection. The event-mode operation of LumaCam distinguishes it from conventional scintillator-based X-ray cameras. Individual scintillation photons are detected and time-stamped, allowing every interaction to be reconstructed in both space and time. Using a scintillator-coupled optical detector with Timepix3 readout, we achieved a spatial resolution of 50~{\textmu}m in X-ray imaging, demonstrating sub-pixel resolution. Thicker scintillators showed only a moderately worse spatial resolution. This is important for high-energy X-ray applications, where thicker scintillators provide significantly higher interaction probabilities. It should be noted that the measured resolution is the intrinsic detector resolution and the sample resolution in a cone-beam configuration can be significantly better, dependent on the setup geometry. The temporal resolution was estimated to range from approximately 15~ns to 1~$\mu$s, depending on the scintillator material and detector configuration. The measured spatial resolution is not considered a fundamental limit of the detector, but is primarily determined by the scintillator thickness and optical configuration, both of which can be further optimized for resolution.

    LumaCam detectors provide an effective combination of high spatial resolution and precise timing, making the detector particularly attractive for studies of dynamic processes. In parallel-beam geometries the detector resolution directly determines the achievable sample resolution. In a cone-beam configuration, a high detector resolution is especially important if a sample cannot be placed close to the X-ray source (e.g. due to sample environments or the sample geometry), and the achievable cone-beam magnification is therefore limited. The lens-coupled architecture with mirrors places sensitive electronics outside the primary radiation beam. It also allows straightforward replacement of the scintillator screen, thereby enabling the detector to be optimized for different applications. Furthermore, the design facilitates X-ray and gamma detection in isolated environments such as vacuum chambers by collecting the scintillation light through a window and placing the complex detector components outside.

    Continuous event acquisition enables quasi-stroboscopic measurements of repetitive processes without requiring pulsed detector operation. Unlike conventional stroboscopic imaging, the radiation source can operate continuously while all detected events are recorded without acquisition gaps or the dead times associated with frame-based readout. Since each event is individually time-stamped, its temporal position within the process cycle can be assigned during offline data analysis. Consequently, long delays between the trigger signal and the process of interest do not reduce the measurement efficiency or require long acquisition windows, providing considerably greater flexibility for investigating dynamic phenomena over a wide range of timescales.

    The comparison of four scintillator materials demonstrates that detector performance can be tailored to specific experimental requirements. Gd$_2$O$_2$S:Ce,Pr,F provides the highest light output and remains well suited for applications that require high detection efficiency. In contrast, single-crystal scintillators, particularly YAG:Ce and GAGG+, offer substantially improved timing performance, while YAP:Ce provides the best estimated timing resolution of about 15~ns due to its fast scintillation decay. The observed dependence of the reconstructed photon multiplicity spectrum on gamma-ray energy, especially for YAG:Ce, also indicates the potential for energy-selective measurements. Although further studies using calibrated photon beams are required to quantify the achievable energy resolution, these results suggest an additional capability beyond high-resolution imaging.

    Future work will focus on optimizing scintillator materials, screen thickness, optical coupling, and reconstruction algorithms to improve the balance between spatial resolution, timing performance, detection efficiency, and count-rate capability. These developments will help identify optimal detector configurations for different energy ranges and applications.

\begin{acknowledgements}
    We would like to thank the Associate Professorship of Food Process Engineering of the Technical University of Munich for providing access to their X-ray device, and especially Dr. Sebastian Gruber for his support in organizing and performing the X-ray measurements. We would also like to thank the Deutsche Forschungsgemeinschaft (DFG, German Research Foundation) who supported the acquisition of the X-ray device under grant number 538007702. Further thanks go to the Bundesministerium für Forschung, Technologie und Raumfahrt (BMFTR, German Federal Ministry of Research, Technology and Space) for their fiancial support in the framework of the research project 05K22WO5  and to the Czech Ministry of Education, Youth and Sports project LM2023040 CERN-CZ.
\end{acknowledgements}

\section*{Competing Interests}
    A. Losko is the director of LoskoVision GmbH, a company developing detector technologies related to those described in this work. A. Wolfertz was employed by LoskoVision GmbH and received salary support from the company during part of this project. O. Zapadlik is a research scientist at CRYTUR, a company developing crystal materials related to this work.  The remaining authors declare no competing interests.

\section*{Data Availability Statement}
    The data used for the research presented in this paper can be made available upon reasonable request.

\section*{Author Contribution}
    \textbf{Alexander Wolfertz}: Conceptualization (supporting); Data Curation (lead); Formal Analysis (equal); Investigation (lead); Methodology (equal); Resources (equal); Software (lead); Original Draft Preparation (lead); Review \& Editing (equal)

    \textbf{Ondrej Zapadlik}: Investigation (supporting); Resources (equal); Original Draft Preparation (supporting); Review \& Editing (equal)

    \textbf{Alex Gustschin}: Data Curation (supporting); Formal Analysis (supporting); Software (supporting); Original Draft Preparation (supporting); Review \& Editing (equal)

    \textbf{Adrian Losko}: Conceptualization (lead); Investigation (supporting); Methodology (equal); Resources (equal); Review \& Editing (equal)

    \textbf{Andrei Nomerotski}: Conceptualization (supporting); Formal Analysis (equal); Investigation (supporting); Methodology (equal); Resources (equal); Software (supporting); Original Draft Preparation (supporting); Review \& Editing (equal)

\bibliography{bibliography}

\end{document}